\documentclass[twocolumn,secnumarabic,amssymb, nobibnotes, aps, prd]{revtex4-2}

\usepackage{CJK} %

\usepackage{amsmath}
\usepackage{graphics}
\usepackage{grffile}
\usepackage{dcolumn}
\usepackage{bm}
\graphicspath{ {./figs/} }
\usepackage{natbib}
\usepackage{nameref}
\usepackage{hyperref}

\begin{document}

\title{Complete 2$\pi$~Phase Control by Photonic Crystal Slabs} 

\author{Mingsen Pan}
\affiliation{Department of Electrical Engineering, University of Texas at Arlington, Arlington, Texas 76019, USA}

\author{Zhonghe Liu}
\affiliation{Department of Electrical Engineering, University of Texas at Arlington, Arlington, Texas 76019, USA}

\author{Akhil Raj Kumar Kalapala}
\affiliation{Department of Electrical Engineering, University of Texas at Arlington, Arlington, Texas 76019, USA}

\author{Yudong Chen}
\affiliation{Department of Electrical Engineering, University of Texas at Arlington, Arlington, Texas 76019, USA}

\author{Yuze Sun}
\email[]{sun@uta.edu}
\affiliation{Department of Electrical Engineering, University of Texas at Arlington, Arlington, Texas 76019, USA}

\author{Weidong Zhou}
\email[]{wzhou@uta.edu}
\affiliation{Department of Electrical Engineering, University of Texas at Arlington, Arlington, Texas 76019, USA}


\email{sun@uta.edu} 
\email{wzhou@uta.edu} 
\date{\today}



\begin{abstract}
Photonic crystal slabs are the state of the art in studies for the light confinement, optical wave modulating and guiding, as well as nonlinear optical response. Previous studies have shown abundant real-world implementations of photonic crystals in planar optics, metamaterials, sensors, and lasers. Here, we report a novel full 2$\pi$ phase control method in the reflected light beam over the interaction with a photonic crystal resonant mode, verified by the temporal coupled-mode analysis and S-parameter simulations. Enhanced by the asymmetric coupling with the output ports, the 2$\pi$ phase shift can be achieved with the silicon photonics platforms such as Silicon-on-Silica and Silicon-on-Insulator heterostructures. Such photonic crystal phase control method provides a general guide in the design of phase-shift metamaterials, suggesting a wide range of applications in the field of sensing, spatial light modulation, and beam steering. 
\end{abstract}

\maketitle 

\section{Introduction}
Phase shift control has been the pursuit across various fields in optics and photonics. The increasing demand for compact integrated devices with higher sensitivity, faster information processing, and more versatile functionalities has driven the development of artificial media (i.e., metamaterials) for decades  \cite{RN13}. Based on the phase mapping paradigm, excellent results have been achieved for holography \cite{RN11, RN10}, metalens \cite{RN14, RN15, RN9}, optical phased array for beam steering \cite{RN16, RN17, RN18}, and phase-based sensors \cite{RN57, RN35, RN58}. The previous efforts have seen the realization of a complete 2$\pi$ phase shift with liquid crystals \cite{RN20, RN19, RN7, RN21}, mechanically reconfigurable metamaterial \cite{RN22}, and electrically modulated cavity resonance \cite{RN24, RN17}. Both the liquid-crystal-assisted and mechanically controlled dielectric metamaterials introduce the 2$\pi$ phase shift from the effective optical path control of the light propagation. Another more compact design is the cavity resonance-induced phase shift in the reflected or transmitted waves.
\par
The phase shift phenomena due to the interaction between light scattering and cavity resonant modes have been reported by multiple studies \cite{RN6, RN17, RN27, RN28, RN29, RN30, RN31, RN32, RN38}. Reference ~\cite{RN17} proposes a phase shifter by interacting the incident light with a Fabry Perot (FP) mode confined within two distributed Bragg reflectors (DBRs). Although this design enables the active control of the phase for the reflected wave, it is still very complicated in the vertical dimension. Reference ~\cite{RN32} reveals that a full 2$\pi$ phase shift can be acquired in an asymmetric 1-dimensional (1D) structure by coupling a defect mode to the transmitted wave. However, the origin and modulation of such phase shift control in the proposed heterostructure is left unexplained. Reference ~\cite{RN38} reports a 2$\pi$ phase shift design with enhanced transmission assisted by Fabry Perot and Mie resonance of a nanopost metasurface. Here, we report a more compact and efficient phase shifter design with a 2-dimensional (2D) photonic crystal slab (PCS) that consists of periodic lattices of air holes on a dielectric layer. The air-hole-based photonic crystal can be integrated for reconfigurable structures with improved thermal and mechanical properties. Guided resonant modes exist in the PCS due to the dielectric permittivity modulation and they reside above the light line and couple to the radiation modes \cite{RN25, RN26, RN33}. Such interaction between radiated wave and guided resonance mode provides an opportunity to achieve a complete 2$\pi$ phase shift and has potential to simplify the complex structure of liquid crystals and DBRs with a thin layer of PCS. 
\par
The theoretical foundation to analyze such mode interaction is built upon the temporal coupled-mode theory, considering the time-reversal symmetry of the PCS structure, and Lorentz reciprocity of the guided resonance \cite{RN3, RN1, RN34, RN2}. We derive the scattering matrix of the 2-port system with the PCS with the temporal coupled-mode analysis. Induced by the asymmetricity of the heterostructure design, the reflected wave features a complete 2$\pi$ phase shift around the PCS guided resonance region. We analyze the condition that favors 2$\pi$ phase shift and propose two types of practical heterostructures that utilize the enhanced reflection by a PCS broadband reflector (BBR) and the FP mode in the dioxide layer of the Silicon-on-Insulator (SOI) device. The phase properties of the PCS structure are numerically simulated using the finite-element method (FEM) by COMSOL Multiphysics solver. The simulation and calculation results show the 2$\pi$ phase shift in the reflected wave can be optimized and manipulated. 

\section{Photonic crystal slab phase shift analysis}

\begin{figure}[htbp]
\center\includegraphics[width=8.6cm]{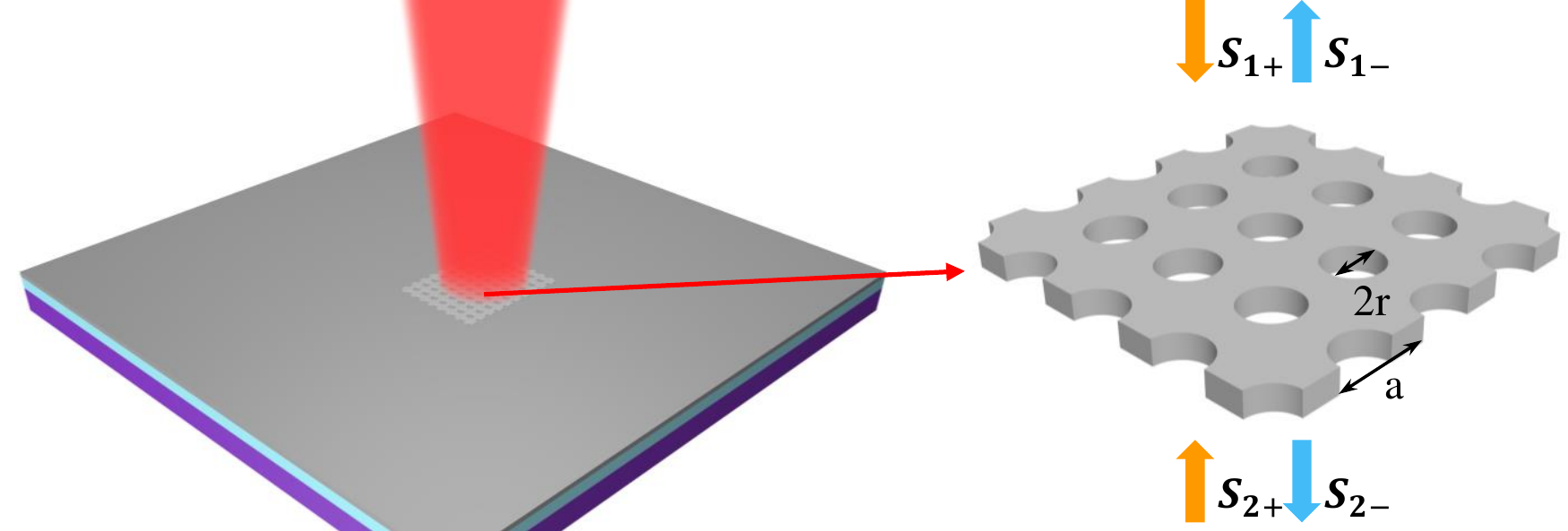}%
\caption{Schematic of the photonic crystal slab on a silicon-on-insulator wafer.}%
\label{fig:fig1}
\end{figure}

Photonic crystal slab (PCS) consists of a periodic array of air holes on the device layer, as shown in Fig. \ref{fig:fig1}. Different optical properties can be acquired by designing the lattice constant (a), air hole radius (r), and the thickness of the PCS. To model the phase shift induced by the PCS resonant modes, we assume a normal incidence where the 90-degree rotational symmetry of the square lattice makes the reflection and transmission spectra polarization-independent. The S-parameters under Ex (or Ey) polarized incidence can be derived from unit-cell simulations with periodic boundary conditions. 

In a 2-port system, the dynamic equations for a resonant mode guided by a PCS structure with an amplitude \textbf{a} read \cite{RN25},
\begin{equation}
\label{eqn:eq1}
\frac{d\textbf{a}}{dt}=(j\omega_0-\frac{1}{\tau})\textbf{a}+\boldsymbol\kappa\textbf{s}_+ ,
\end{equation}
\begin{equation}
\label{eqn:eq2}
\textbf{s}_-=\textbf{C}\textbf{s}_++\textbf{a}\textbf{d}
\end{equation}
where $\textbf{s}_+=[s_{1+}~~s_{2+}]^\textbf{T}$ and $\textbf{s}_-=[s_{1-}~~s_{2-}]^\textbf{T}$ are the input and output field amplitudes to the two ports, respectively. The PCS resonant mode couples with incoming waves $\textbf{s}_+$ with coupling constants $\boldsymbol\kappa=[\kappa_1~~\kappa_2]$ and the outgoing waves $\textbf{s}_-$ with coefficients $\textbf{d}=[d_1~~d_2]^\textbf{T}$. Eq. \eqref{eqn:eq1} and Eq. \eqref{eqn:eq2} describe the temporal coupling for mode \textbf{a} with resonant frequency $\omega_0$ and life time $1/\tau$ with the background scattering matrix \textbf{C}. Determined by the constraints set by the scattering matrix unity and time-reversal symmetry, the total scattering matrix can be derived as
\begin{equation}
\label{eqn:eq3}
\textbf{S}=\textbf{C}+\frac{\textbf{d}\textbf{d}^\textbf{T}}{j(\boldsymbol\omega-\omega_0)+1/\tau}.
\end{equation}
\par
The first term in the complex-valued scattering matrix \textbf{S} represents the background scattering of light waves, while the second term shows the resonant mode interacting with the background scattering via the coupling term \textbf{d}. 
\subsection{Symmetric structures}
It is well known that 2$\pi$ phase shift is impossible in a vertically symmetric PCS structure where $|d_1|^2=|d_2|^2=1/\tau$. The scattering matrix of this symmetric structure can be further simplified, considering a special form of background scattering matrix \cite{RN34} $\textbf{C}=exp(j\theta)\begin{bmatrix}
r & jt\\
jt & r
\end{bmatrix}$, 
\begin{equation}
\label{eqn:eq4}
\textbf{S}_\textbf{11}(\boldsymbol\omega)=exp(j\theta)\{r-\frac{1/\tau}{j(\boldsymbol\omega-\omega_0)+1/\tau}(r \pm jt)\}
\end{equation}
where the $\pm$ sign denotes the parity of the resonant mode. It can be deduced that $\textbf{S}_\textbf{11}(\boldsymbol\omega)$ evolves with the incoming wave frequency and passes 0 point on its complex plane when $\boldsymbol\omega=\omega_0 \pm \frac{t}{r}(1/\tau)$. The phase shift cannot cover the full $2\pi$ range in the reflection spectrum because the phase singularity creates a discontinuity or a sudden jump in the reflection phase. At the phase singularity point, the reflection drops to 0 due to the destructive interference between the resonant mode and the background reflection. Therefore, a symmetric structure does not support a 2$\pi$ phase shift due to the singularities in the reflection or transmission spectra. 
\par

\begin{figure*}[htbp]
\center\includegraphics[width=12.3cm]{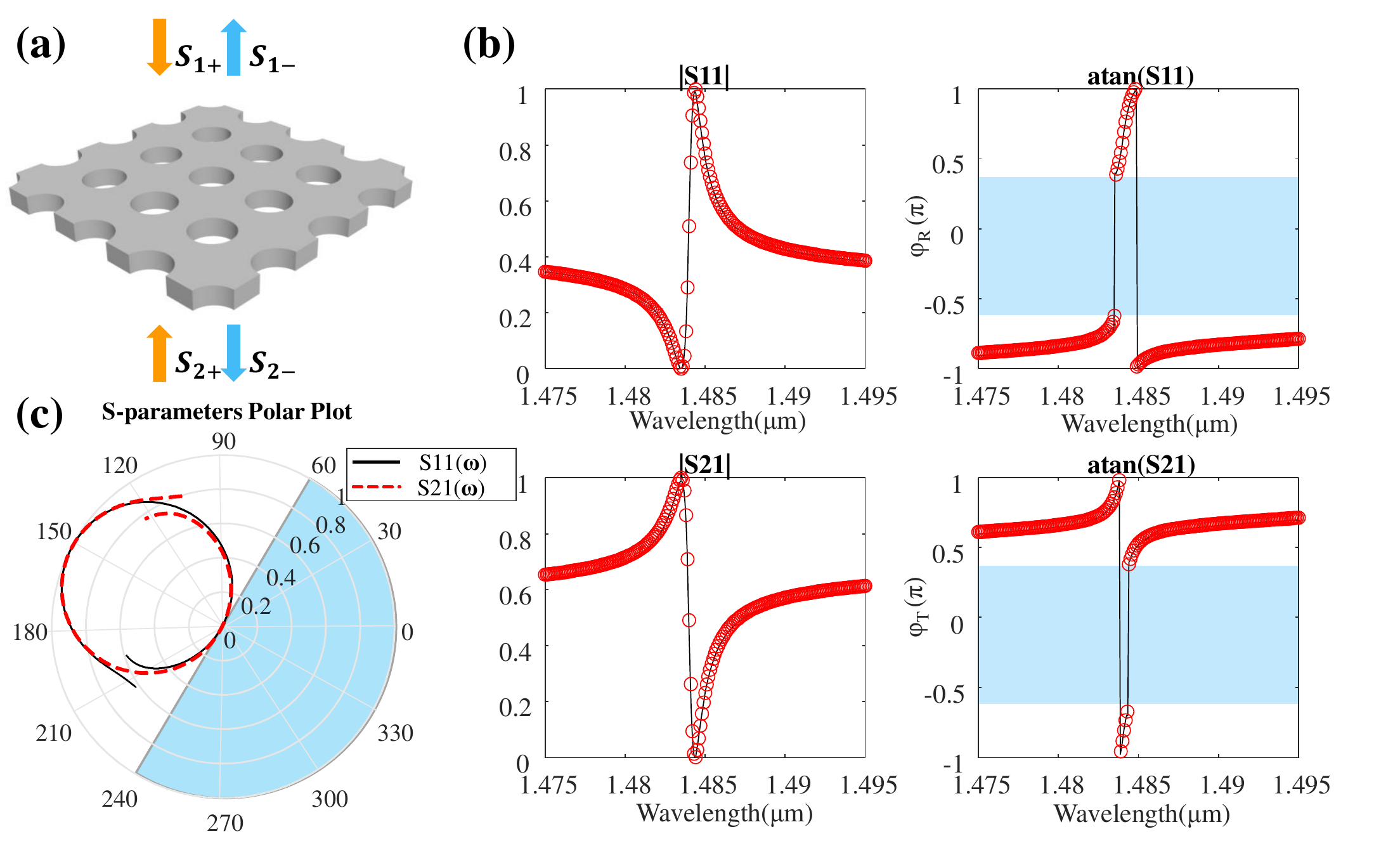}%
\caption{Reflectance and phase shift spectrum for a vertically symmetric photonic crystal slab. \protect\hypertarget{fig2a}{(a)} Definition of the symmetric PCS two-port scattering model. The PC layer is formed by square lattice with one circular air hole in each unit cell. The Silicon layer index is 3.4858 surrounded by air. \protect\hypertarget{fig2b}{(b)} Reflection and transmission spectra derived from the scattering matrix $\textbf{S}(\boldsymbol\omega)$ (solid black lines) and simulated using FEM (red circles). The phase shifts in the shaded regions are undefined due to the phase singularities. \protect\hypertarget{fig2c}{(c)} Polar plots of $S_{11}(\boldsymbol\omega)$ and $S_{21}(\boldsymbol\omega)$ evolved within spectral range from $1.475\mu m$ to $1.495\mu m$. The plots pass the zero point, and the shaded region shows zero reflection around the resonance frequency, thus creating the abrupt phase shift of $\pi$. }%
\label{fig:fig2}
\end{figure*}
Let us take the PCS in the air as an example for a symmetric system. We consider the PCS structure has a thickness of 250nm, with lattice constant $a=1\mu$m, and air filling ratio r/a = 0.08, shown in Fig. \ref{fig:fig2}\hyperlink{fig2a}{(a)}. The background scattering and mode coupling parameters are simulated with finite-element method (FEM) by COMSOL Multiphysics. The PCS has a guided resonance at $\omega_0=201.98THz$ with lifetime $1/\tau=0.0570THz$ that couples to the $\textbf{Ex}$-polarized incoming wave. Using the coupled-mode theory mentioned in the main text, the mode coupling constants to the two ports are $|d_1|^2=|d_2|^2=0.0143\%$. The scattering matrix characterizes the relations between the incoming and outgoing states, such as the reflectance and reflection phase shift spectra shown in the top panel in Fig. \ref{fig:fig2}\hyperlink{fig2b}{(b)}, and the transmittance and transmission phase change shown in the bottom panel in Fig. \ref{fig:fig2}\hyperlink{fig2b}{(b)}. As can be seen from the spectra, we have a reflection zero at $\omega=1.4835\mu m (202.085THz)$ and a transmission zero at $\omega=1.484\mu m (201.97THz)$ where $S11(\boldsymbol\omega)$ and $S21(\boldsymbol\omega)$ pass zero, respectively, creating the $\pi$ jump in the phase shift spectrum (Fig. \ref{fig:fig2}\hyperlink{fig2c}{(c)}). Therefore, in a symmetric system, we cannot achieve a continuous 2$\pi$ phase shift due to the existence of the phase singularities. 

\subsection{Heterostructures}
To pursue a 2$\pi$ phase shift, it is necessary to remove the phase singularities (i.e., reflection zero or transmission zero) by the vertical design, such as the heterostructures. Without considering the special form in Eq. \eqref{eqn:eq4}, the reflection and transmission phase change derived by the $\textbf{S}_{11}$ and $\textbf{S}_{21}$ parameters are 
\begin{equation}
\label{eqn:eq5}
\textbf{S}_{11}(\boldsymbol\omega)=\textbf{C}_{11}+h(\boldsymbol\omega)|d_1|^2 ,
\end{equation}
\begin{equation}
\label{eqn:eq6}
\textbf{S}_{21}(\boldsymbol\omega)=\textbf{C}_{21}+h(\boldsymbol\omega)d_2^*d_1
\end{equation}
where $h(\boldsymbol\omega)=\frac{1}{j(\boldsymbol\omega-\omega_0)+1/\tau}$ is the lineshape function of the PCS resonant mode. With the complex analysis of the S-parameters, a sufficient condition for 2$\pi$ phase change in reflected wave can be derived as,
\begin{equation}
\label{eqn:eq7}
|d_1|^2>|r|/|h(\boldsymbol\omega)|.
\end{equation}
where a 2$\pi$ phase shift for the reflected wave can be achieved in a structure that has stronger mode coupling amplitude $d_1$. In a symmetric 2-port system, the mode profile decays equally to the two ports and the coupling constants are $|d_1|^2=|d_2|^2=1/\tau$. However, for a heterostructure, the mode extends to the two ports asymmetrically. To describe such asymmetric coupling under the energy conservation constraint, we define a reflection coupling enhancement factor as
\begin{equation}
\label{eqn:eq8}
\Gamma=\frac{|d_1|^2}{|d_2|^2}.
\end{equation}
\par
The factor shows the asymmetricity of outgoing energy of the mode to the two ports of the heterostructure. 
Eq. \eqref{eqn:eq5} and \eqref{eqn:eq6} neglects the slow variation of the background scattering and assumes a uniform background scattering at  $\omega =\omega _0$. The reflectance \textit{R} and the phase shift in reflected wave 
$\varphi _R$ \ can be obtained through the intensity and phase of the  $\text S_{11}\left(\boldsymbol\omega \right)$ \ term.
\begin{eqnarray}
\label{eqn:eq9}
R\left(\omega \right)=\left|\text S_{11}\left(\boldsymbol\omega \right)\right|^2,
\\
\label{eqn:eq10}
\varphi _R\left(\omega \right)=\mathit{atan}(\text S_{11}\left(\boldsymbol\omega \right)).
\end{eqnarray}

\begin{figure}[htbp]
\center\includegraphics[width=7cm]{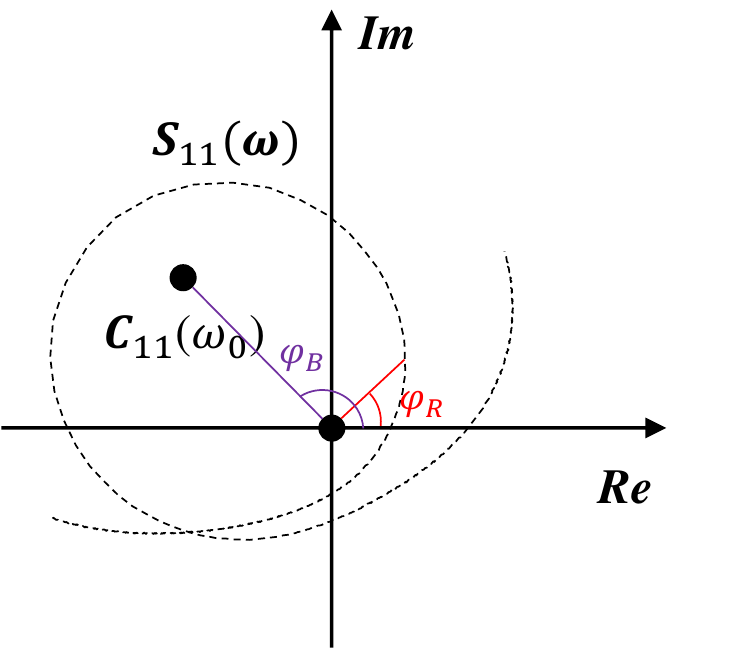}%
\caption{The  $\text S_{11}\left(\boldsymbol\omega \right)$ \ \ parameter on the complex plane. $\varphi_B$ is the background reflection phase shift and $\varphi_R$ is the total reflection phase shift with the PCS. When the $\text S_{11}\left(\boldsymbol\omega \right)$ curve encircles the zero point witin a spectral range, the phase shift is a complete $2\pi$.}%
\label{fig:fig3}
\end{figure}

\par
From the complex analysis of the scattering matrix, the condition for 2$\pi $ phase shift is that the S-parameters encircle the 0 point in their complex plane when they are evolving with the frequency  $\boldsymbol\omega $. The second term $h\left(\omega \right)\left|d_1\right|^2$ \ determines the phase delay that the incident wave experiences when reflected away from the PCS. It depends on two factors, the spectral lineshape  $h\left(\omega \right)$ \ and the mode coupling amplitude  $\left|d_1\right|$. As shown in Fig. \ref{fig:fig3}, by increasing  $\left|d_1\right|^2$, the amplitude of $h\left(\omega \right)\left(d_1\right)^2$ \ can be enlarged and the condition to encircle the 0 point is more favorable to satisfy. On the other hand, the  $\left|d_1\right|^2$ \ cannot be enhanced infinitely due to the energy conservation, 
\begin{equation}
\label{eqn:eq11}
\left|d_1\right|^2+\left|d_2\right|^2=\frac 2{\tau }, 
\end{equation}
which puts another constraint on the phase shift condition. Through the coupling with the resonant mode, the reflected or transmitted wave acquires an additional phase shift. The sufficient condition for a 2$\pi $ phase shift in reflected wave is that the amplitude of  $h\left(\boldsymbol\omega \right)\left(d_1\right)^2$ \ is larger than that of the background reflection  $\left|C_{11}\left(\boldsymbol\omega
\right)\right|$. Therefore, the mode coupling  $\left|d_1\right|^2$ \ to achieve 2$\pi $ reflection phase shift satisfies
\begin{equation}
\label{eqn:eq12}
\left|C_{11}\left(\boldsymbol\omega \right)\right|\ast \sqrt{\left(\boldsymbol\omega -\omega _0\right)^2+\left(\frac 1{\tau
}\right)^2}<\left|d_1\right|^2<\frac 2{\tau }
\end{equation}
\par
Substituting $\left|d_1\right|^2=\left|d_2\right|^2=\frac 1{\tau }$ into Eq. \eqref{eqn:eq12}, we can derive the 2$\pi $ phase shift condition in a vertically symmetric structure, which is,
\begin{equation}
\label{eqn:eq13}
r<\frac{1/\tau }{\sqrt{\left(\boldsymbol\omega -\omega _0\right)^2+\left(1/\tau \right)^2}}
\end{equation}
where  $r=\left|C_{11}\left(\boldsymbol\omega \right)\right|$. The right side in Eq. \eqref{eqn:eq13} is no larger than 1. However, there exists an extreme condition where Eq. \eqref{eqn:eq13} is not satisfied, 
\begin{equation}
\label{eqn:eq14}
\boldsymbol\omega=\omega_0 \pm \frac{t}{r}(1/\tau)
\end{equation}
which makes  $r=\frac{1/\tau }{\sqrt{\left(\boldsymbol\omega -\omega _0\right)^2+\left(1/\tau \right)^2}}$. In Eq. \eqref{eqn:eq14}, \textit{t} is the transmission coefficient derived from the energy conservation,  $\left|r\right|^2+\left|t\right|^2=1$. The condition of Eq. \eqref{eqn:eq14} matches the results derived from Eq. \eqref{eqn:eq4} where the polar plot of  $\text S_{11}\left(\boldsymbol\omega \right)$ \ shows it is a singularity that causes a $\pi $ phase shift jump.
In this regard, we can further derive a general condition for $2\pi$ phase shift in reflection as $\Gamma>1$ or $|d_1 |^2>1/\tau>|d_2 |^2$. To engineer a robust phase shift device, the objective function to optimize the system to achieve a $2\pi$ phase shift is hence to maximize $\Gamma$.
On the other hand, \eqref{eqn:eq12} indicates that a suitable Q-factor of the resonant mode is also necessary. The Q-factor can be calculated by the 
\begin{equation}
\label{eqn:eq15}
{\Delta}\omega ^2+(\frac 1{\tau })^2<\frac{\left|d_1\right|^4}{\left|r\right|^2}
\end{equation}
where  ${\Delta}\omega =(\boldsymbol\omega -\omega _0)$ \ is the frequency range for the phase shift. Divided by  $\omega _0^2$,
\begin{equation}
\label{eqn:eq16}
\frac{{\Delta}\omega ^2}{\omega _0^2}+(\frac 1{2Q})^2<\frac{\left|d_1\right|^4}{\omega_0^2\left|r\right|^2}
\end{equation}
where  $2Q=\omega _0\tau $. Finally, we reach an inequality for the Q-factor,
\begin{equation}
\label{eqn:eq17}
Q>\frac 1 2\left(\frac{\left|d_1\right|^4}{\omega _0^2\left|r\right|^2}-\frac{{\Delta}\omega ^2}{\omega_0^2}\right)^{-1/2}
\end{equation}
\par
Considering Eq. \eqref{eqn:eq8} and Eq. \eqref{eqn:eq11}, we can get another condition for the 2$\pi $ phase shift, 
\begin{equation}
\label{eqn:eq18}
\left|d_1\right|^2>\frac 1{\tau },
\end{equation}
and substituting into Eq. \eqref{eqn:eq15} and using  $2Q=\omega _0\tau $, we can derive the upper boundary for Q-factor, 
\begin{equation}
\label{eqn:eq19}
Q<\frac 1 2\frac{\left|t\right|}{\left|r\right|}\frac{\omega _0}{\left|{\Delta}\omega\right|}. 
\end{equation}
\par
Eq. \eqref{eqn:eq17} indicates that higher quality factor helps with the phase shift conditions because the faster the phase shift happens, the less possibility for the mode to couple with other modes. Eq. \eqref{eqn:eq19} reveals that if the Q-factor is too large, it will not couple very well to the outgoing and incoming waves to perform the phase shift. In our PCS design, thanks to its appearance above the light line, the PCS guided resonance modes has lower Q-factor compared to the bandgap cavity modes, mostly in the order of 10\textsuperscript{3}. Our simulation results show that several thousand of Q-factor for the PCS modes will satisfy Eq. \eqref{eqn:eq18} where the coupled energy to the reflected wave should be larger than half the energy stored in the cavity as the guided mode resonance. Therefore, the tuning of  $\left|d_1\right|^2$ \ is more sensitive and effective to control the phase shift condition. In conclusion, the requirements for the mode to introduce the 2$\pi $ phase shift are that it has relatively higher Q-factor than the background, and its enhancement factor  $\Gamma >1$ \ to avoid the singularity. 

\subsection{Calculation of reflection enhancement factor}
To calculate the reflection enhancement factor defined in Eq. \eqref{eqn:eq8}, we derive from the scattering matrices \textbf{S} and
\textbf{C }at the resonant position $\boldsymbol\omega =\omega _0$\textbf{ }, 
\begin{eqnarray}
\label{eqn:eq20}
\text S_{11}\left(\omega _0\right)=\text C_{11}(\omega _0)+\tau \left(d_1\right)^2,
\\
\label{eqn:eq21}
\text S_{22}\left(\omega _0\right)=\text C_{22}(\omega _0)+\tau \left(d_2\right)^2. 
\end{eqnarray}
The reflection enhancement factor is thus, 
\begin{equation}
\label{eqn:eq22}
\Gamma =\left|\frac{S_{11}\left(\omega _0\right)-\text C_{11}(\omega _0)}{\text S_{22}\left(\omega _0\right)-\text
C_{22}(\omega _0)}\right|
\end{equation}
\par
The S $(\omega _0)$ parameters are obtained numerically from the reflection spectra of the PCS structure, and the background $C(\omega _0)$ parameters can be approximated by fitting the slowly varying background from the spectra. To get better fitting results, we can adopt the background fitting method by using the effective medium of the PCS layer. The optical properties of the PCS is approximated by a uniform layer of dielectric material whose refractive index can be effectively calculated by the relation $\varepsilon _{eff}=(1-f_1)\varepsilon _{Si}+f_1\varepsilon _{air}$ \ where the filling ratio is $f_1=\pi r^2/a^2$. 
\par
Another method to calculate the enhancement factor is to retrieve the field profiles of the eigenmodes. Assuming the incoming wave amplitudes from the two ports are zero (i.e., no incoming waves), the field of the PCS resonant mode will decay exponentially from the PCS layer to the two ports. The outgoing wave amplitude is stronger at the port that has stronger mode coupling. The ratio of the wave intensities at the end of the two ports is the enhancement factor according to the definition of Eq. \eqref{eqn:eq8}. The energy density distribution can be calculated from the surface integration at each z plane, 
\begin{equation}
\label{eqn:eq23}
W(z)=\iint _{z=a}\mathcal{E}(x,y)\mathit{dxdy}
\end{equation}
where the energy density  $\mathcal{E}(x,y)$ \  is given by
\begin{equation}
\label{eqn:eq24}
\mathcal{E}\left(x,y\right)=\varepsilon \left|E(x,y)\right|^2
\end{equation}
where  $\varepsilon $ \ is the relative dielectric permittivity and  $E(x,y)$ \ is the electric field amplitude at the point (x, y, a). Assuming no incoming power, the power ratio of the outgoing wave to one port and the total energy trapped inside the PCS is the mode coupling strength to that port. Therefore, the enhancement factor  $\Gamma=\frac{\left|d_1\right|^2}{\left|d_2\right|^2}$ \ can be calculated as

\begin{equation}
\label{eqn:eq25}
\Gamma =\frac{\iint _{z=\mathit{b1}}\mathcal{E}(x,y)\mathit{dxdy}}{\iint _{z=\mathit{b2}}\mathcal{E}(x,y)\mathit{dxdy}}
\end{equation}
where b1 and b2 are integration planes near port 1 and port 2, respectively.

\section{Enhanced mode coupling}
The last section has shown that the asymmetric structure, with enhanced coupling factor $\Gamma$, can help achieve full 2$\pi$ reflection phase control. To improve the mode coupling to the reflection wave, we consider two types of heterostructures, the compact PCS broadband reflector (silicon on a glass substrate) and the Silicon-on-Insulator (SOI) heterostructure. 
\subsection{PCS broadband reflector}
By tuning the PCS lattice parameters, either narrow-band Fano resonance filters or BBRs can be implemented. The Q-factor of the guided resonance modes plays an import role in determining the spectral lineshape of the resonances \cite{RN26}. Larger air hole filling ratio increases the air-mode coupling for some modes and creates the low-Q broadband lineshape. Such BBRs have high reflectance with a large spectral range and have the promise to enhance mode coupling to the reflected wave. 

\begin{figure}[htbp]
\center\includegraphics[width=8.6cm]{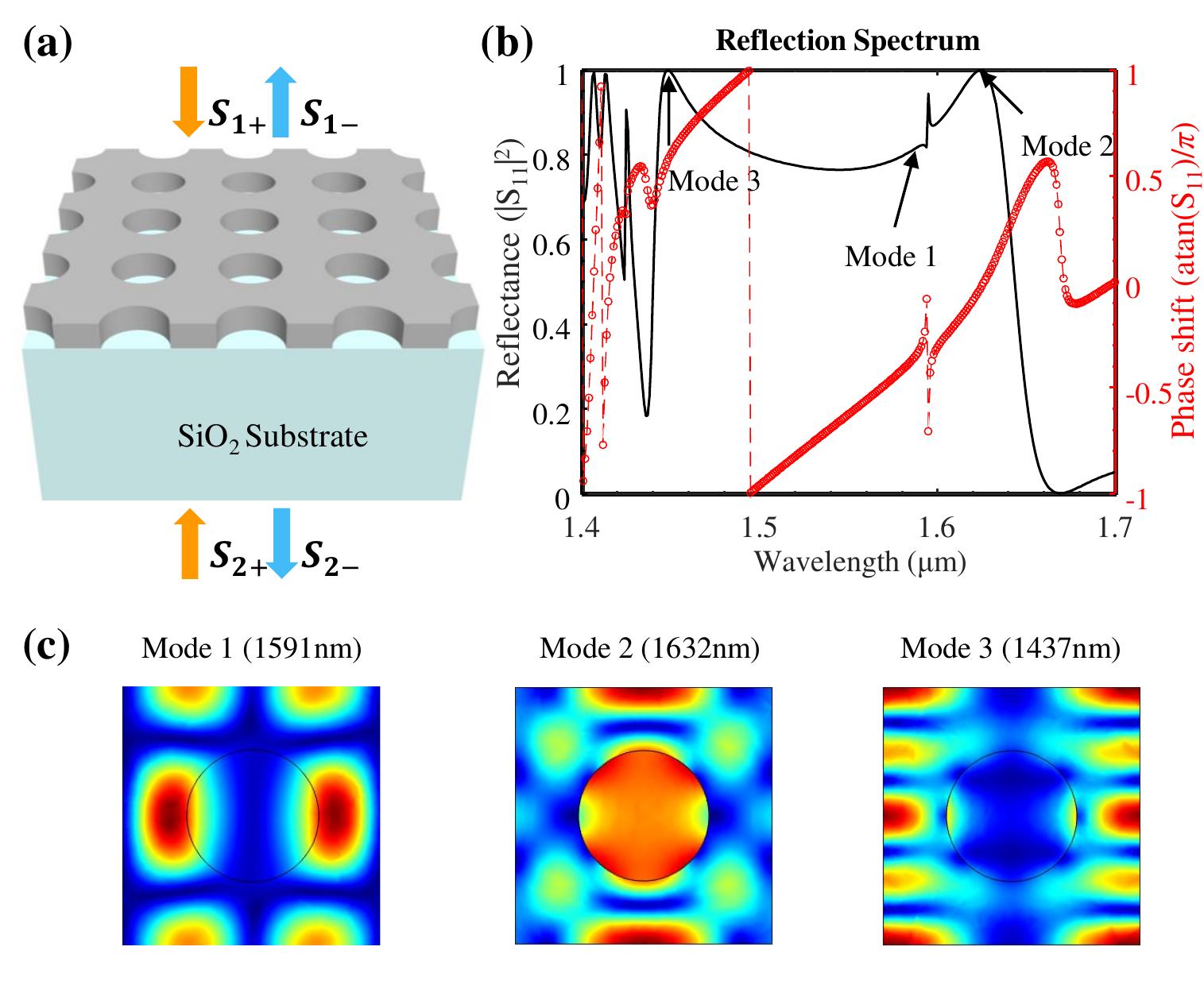}%
\caption{PCS broadband reflector on a glass substrate. \protect\hypertarget{fig4a}{(a)} Schematic of the PCS on a silica substrate. The PCS has lattice constant $a=0.99\mu m$, air filling ratio $r/a=0.25$,  and thickness $h=300nm$. \protect\hypertarget{fig4b}{(b)} Reflection spectra of the BBR derived from S-parameter simlations. The black solid line is the reflectance and the red line with circles is the phase shift in the reflected wave. Mode 2 (208.29THz) and Mode 3 (182.63THz) form the lineshape of the broadband reflection. Mode 1 located in the middle of the BBR has higher Q for inducing the 2$\pi$ phase shift. \protect\hypertarget{fig4c}{(c)} \textbf{|E|} field distribution of the three modes. }%
\label{fig:fig4}
\end{figure}
\par
As shown in Fig. \ref{fig:fig4}\hyperlink{fig4a}{(a)} and Fig. \ref{fig:fig4}\hyperlink{fig4b}{(b)}, the designed BBR sits on a glass substrate and features high-reflection within a broad spectral range from $1.45\mu m$ to $1.62\mu m$. The broadband reflection is formed by the coupling of two low-Q modes (mode 2 and mode 3 as indicated by Fig. \ref{fig:fig4}\hyperlink{fig4b}{(b)}. Simulated by the COMSOL eigenfrequency solver, Mode 2 has a complex eigenfrequency $\omega_2=208.29+j0.6858 THz$ with $Q_2=142.5$, and $\omega_3=182.63+j2.1753 THz$ with $Q_3=42$ for Mode 3. Such structure also supports higher-Q mode with lower-order mode profiles, shown in \ref{fig:fig4}\hyperlink{fig4c}{(c)}, such as Mode 1 located within the BBR spectrum. At guided resonance frequency of $\omega_1=188 THz$, the lifetime and Q-factor for Mode 1 are $1/\tau=0.028THz$ and $Q_1=3311$, respectively. Coupled with the strong background scattering of the BBR, Mode 1 is expected to have enhanced coupling to the reflected wave and hence achieve complete 2$\pi$ phase control. 
\par
\begin{figure*}[htbp]
\center\includegraphics[width=12.3cm]{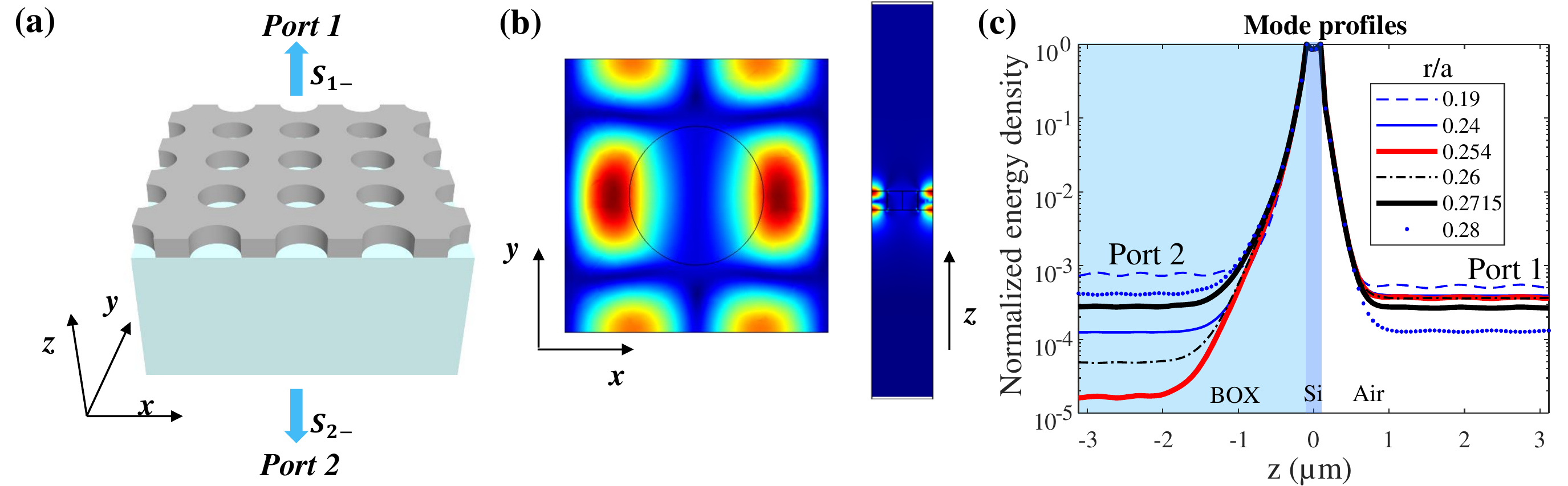}%
\caption{Eigenmode field distribution and reflection enhancement factor calculation. \protect\hypertarget{fig5a}{(a)} The FEM eigenfrequency simulations assumes the guide modes in PCS couple without incoming waves from the two ports. \protect\hypertarget{fig5b}{(b)} The normalized electric field distribution of the PCS guided mode in a unit cell with r/a = 0.254. \protect\hypertarget{fig5c}{(c)} Energy density distribution of the eigenmodes on the PCS structure with air filling ratio r/a = 0.19, 0.24, 0.254, 0.26, 0.2715, and 0.28. }%
\label{fig:fig5}
\end{figure*}
\begin{figure*}[htbp]
\center\includegraphics[width=12.9cm]{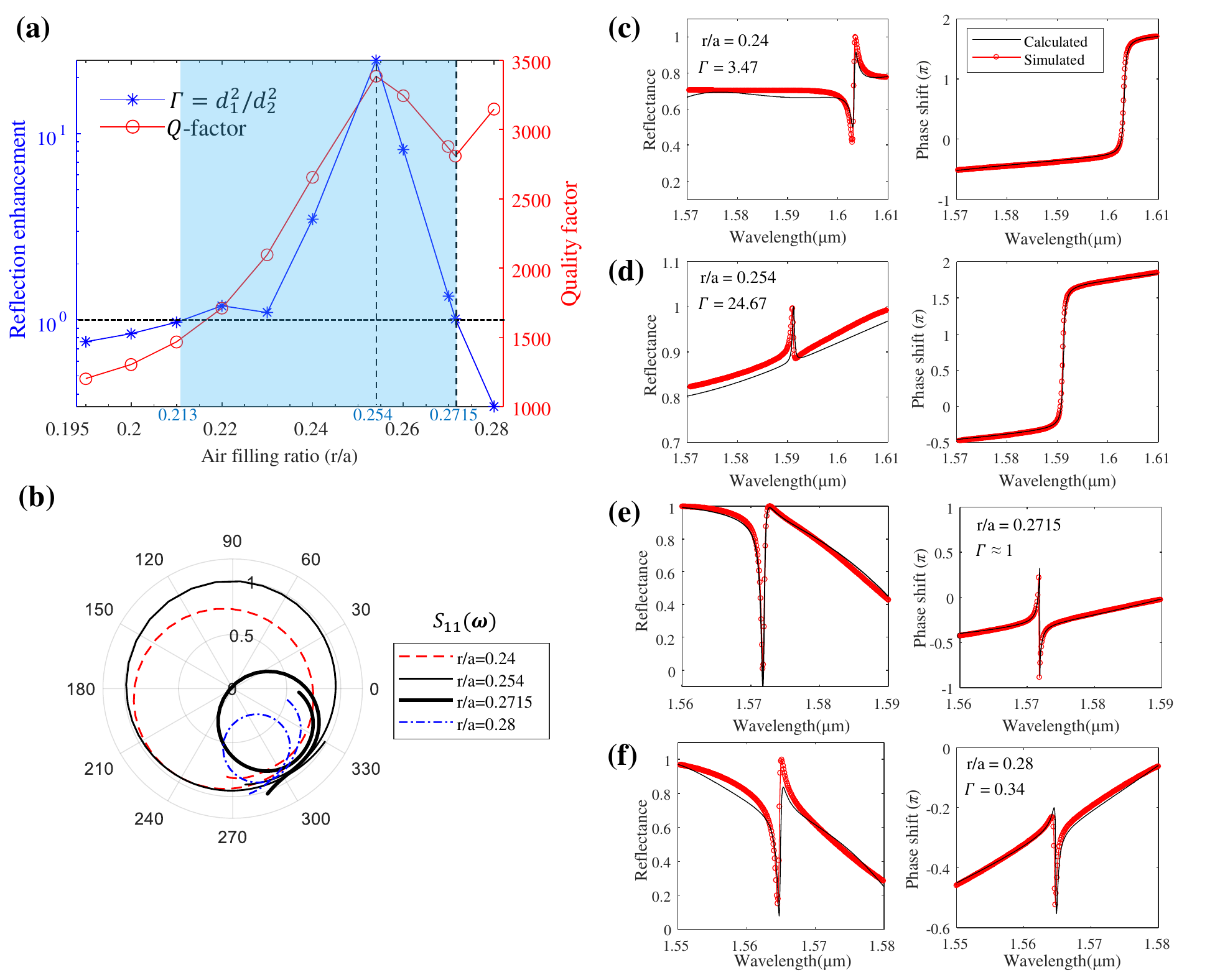}%
\caption{Reflection enhancement factor ($\Gamma$) and phase shift relation. \protect\hypertarget{fig6a}{(a)} $\Gamma$ and Q-factor with different air filling ratios where the shaded region supports complete 2$\pi$ phase control. \protect\hypertarget{fig6b}{(b)} Polar plots of complex $S_{11}(\boldsymbol\omega)$ with four different air filling ratios, r/a=0.24 (red dashed line), 0.254 (thin black solid line), 0.2715 (thick black solid line), and 0.28 (blue dash-dot line). \protect\hypertarget{fig6c}{(c)}-\protect\hypertarget{fig6f}{(f)} Corresponding reflection spectra for the four parameters. Two extreme conditions are observed where $\Gamma$ becomes the strongest \protect\hypertarget{fig6d}{(d)} and $\Gamma\approx1$ \protect\hypertarget{fig6e}{(e)}. }%
\label{fig:fig6}
\end{figure*}

The reflection phase shift within a spectral range is governed by the trace of the $S_{11}(\boldsymbol\omega)$ on its complex plane. A full 2$\pi$ phase shift can happen when the $S_{11}$ curve encircles the 0 point. By utilizing the reflection enhancement factor defined in Eq. \eqref{eqn:eq8}, we can determine if the reflected wave carries 2$\pi$ phase shift within the spectral range. Fig. \ref{fig:fig5}\hyperlink{fig5c}{(c)} shows the vertical field energy density distribution for the PCS modes (Mode 1) calculated by the COMSOL eigenfrequency simulations. There appear several critical points when sweeping the air filling ratio, r/a. When r/a = 0.254, the mode energy profile shows extreme asymmetricity and the energy going towards port 1 is over 20 times larger than that to port 2. When r/a = 0.2715, the energy density becomes very closed to that for a symmetric structure, though the structure is asymmetric. When r/a = 0.19 or r/a = 0.28, fewer energy is coupled into the port 1 and the reflection spectra are not expected to see 2$\pi $ reflection phase shift. Shown in Fig. \ref{fig:fig6}\hyperlink{fig6a}{(a)} are the reflection enhancement factors $\Gamma$ and Q-factors calculated for several different air filling ratios for the PCS structure in Fig. \ref{fig:fig4}\hyperlink{fig4a}{(a)} using the method in Eq. \eqref{eqn:eq23} - Eq. \eqref{eqn:eq25}. Thanks to the vertically asymmetric refractive index above and below the PCS, by tuning the air filling ratio, we can directly control the vertical field profile of Mode 1, thus creating either enhanced or weakened coupling to the reflected wave. The reflection enhancement factor has its peak value of $\Gamma$~=~24.67 when r/a~=~0.254. The trajectories of $S_{11}(\boldsymbol\omega)$ for several different r/a ratios, 0.24, 0.254, 0.2715, and 0.28, shown in Fig. \ref{fig:fig6}\hyperlink{fig6b}{(b)}, indicate the phase shift for the first two parameters have full 2$\pi$ phase control, while the phase shift for r/a~=~0.28 is smaller than $\pi$. This is verified by the coupled-mode calculation and FEM simulations, as shown in Fig. \ref{fig:fig6}\hyperlink{fig6c}{(c)}$-$\hyperlink{fig6f}{(f)}. 

\par 
Different from the previous symmetric case, with a larger reflection coupling strength $|d_1^2|$, the reflected wave does not have reflection zero if we engineer the vertical asymmetricity of the structure. The shaded region in Fig. \ref{fig:fig6}\hyperlink{fig6a}{(a)} where the air filling ratio 0.213 < r/a < 0.2715 has the $\Gamma$ factor larger than 1 and the 2$\pi$ phase shift control can be realized in the reflected wave. The region corresponds to over 50nm air hole radius tuning range, which gives sufficient tolerance for fabrication. At the strongest $\Gamma$, the trace of $S_{11}(\boldsymbol\omega)$ encircles the 0 point in the complex plane and has a uniformly high reflection around the resonance mode during the 2$\pi$ phase shift (Fig. \ref{fig:fig6}\hyperlink{fig6c}{(c)}). As the $\Gamma$ becomes smaller, the reflection spectra become more asymmetric with dips gradually approaching to 0. In theory, we can achieve symmetric coupling strengths ($|d_1^2|=|d_2^2|$ and $\Gamma$=1) even in this asymmetric structure, which needs fine tuning in the parameter spaces. Tuning the r/a ratio to 0.2715 when $\Gamma$ approaches one, the reflection spectrum in Fig. \ref{fig:fig6}\hyperlink{fig6e}{(e)} shows a reflectionless point near 1572nm and there appears an abrupt phase shift around that point. When $\Gamma$ is smaller than 1, the resonant mode will experience stronger coupling to the transmission port than the reflection side, and the phase shift will be smaller than $\pi$, as shown in Fig.\ref{fig:fig6}\hyperlink{fig6f}{(f)}. It is important to note that the mode Q-factor also plays an important role to the phase shift phenomenon. It is obvious that higher Q-factor will have faster transition to the phase shift and separate the mode from the influence of the other background modes. On the contrary, low-Q modes have slow phase shift variations, such as that for Mode 2 and Mode 3, shown in Fig. \ref{fig:fig4}\hyperlink{fig4b}{(b)}, which is not sufficient for $2\pi$ phase shift. Otherwise, the coupling enhancement factor $\Gamma$ has the full control over the behaviors of reflection spectrum, including reflection lineshape and phase shift controls. 

\begin{figure*}[htbp]
\center\includegraphics[width=12.9cm]{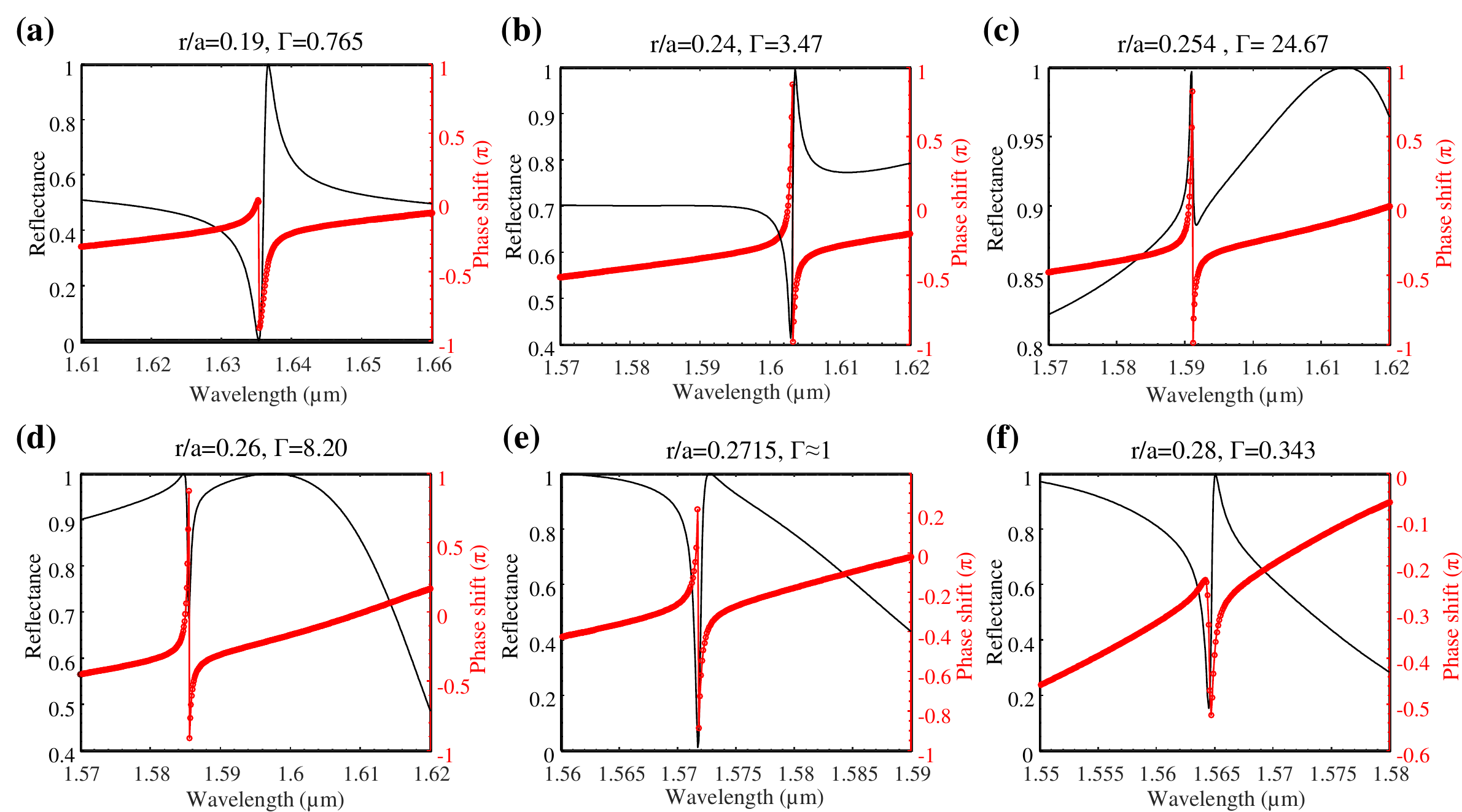}%
\caption{Broadband reflector lineshape control. Reflection lineshape variation with several characteristic air hole filling ratios, i.e., r/a = 0.19 \protect\hypertarget{fig7a}{(a)}, 0.24 \protect\hypertarget{fig7b}{(b)}, 0.254 \protect\hypertarget{fig7c}{(c)}, 0.26 \protect\hypertarget{fig7d}{(d)}, 0.2715 \protect\hypertarget{fig7e}{(e)}, and 0.28 \protect\hypertarget{fig7f}{(f)}. 2$\pi $ phase shift can be observed in (b), (c), and (d). The parity of Fano resonance changes at the critical points at (c) and (e).}%
\label{fig:fig7}
\end{figure*}

\par
The lineshape changes with different air filling ratios, including the parity and asymmetric factor of the Fano lineshape. Fig. \ref{fig:fig7} captures these transitions whose boundaries are marked by two critical conditions. The first critical point is at r/a = 0.254 (Fig. \ref{fig:fig7}\hyperlink{fig7c}{(c)}) where the reflection enhancement ($\Gamma$) is the strongest, and the second is r/a = 0.2715 (Fig. \ref{fig:fig7}\hyperlink{fig7e}{(e)}) where the enhancement factor $\Gamma{\approx}$ 1. This extreme condition of $\Gamma=1$ is very sensitive and fragile. As $\Gamma$ is enhanced by the asymmetric structure, the 2$\pi$ phase shift becomes more stable. When $\Gamma$ becomes smaller than 1, the total phase shift triggered by the PCS mode decreases below $\pi$. The lineshape is determined by the interference between the PCS guided mode and the two low-Q background modes that form the BBR. Interacting with this complex background will give rise to the switch between different parity and Fano asymmetric factor.

\subsection{Silicon-on-insulator heterostructure}
\begin{figure}[htbp]
\center\includegraphics[width=8.6cm]{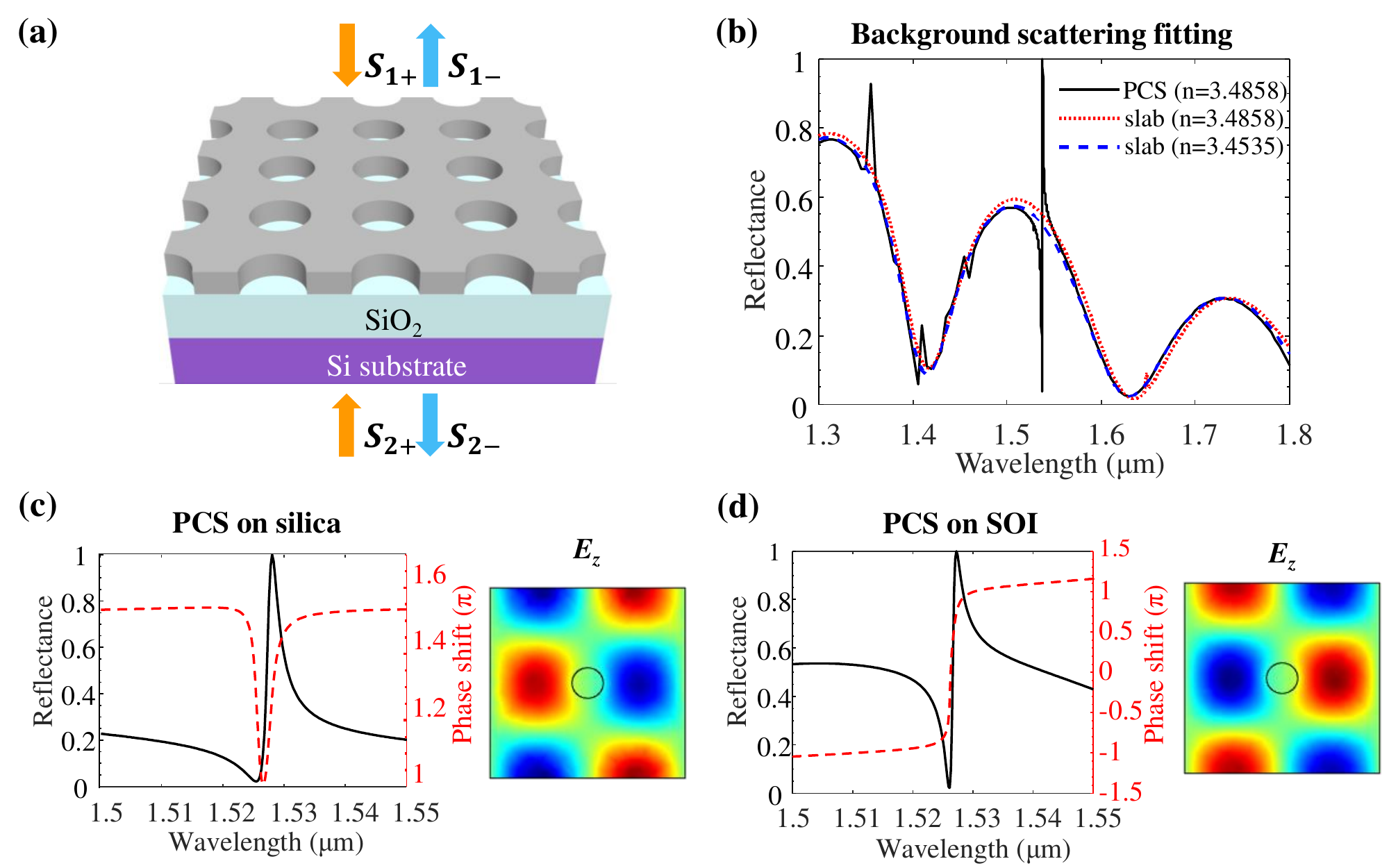}%
\caption{PCS on silicon-on-insulator platform. \protect\hypertarget{fig8a}{(a)} Schematic of the PCS designed on an SOI platform. \protect\hypertarget{fig8b}{(b)} Background scattering fitting of the PCS reflection spectrum. The lattice constant, air filling ratio, and thickness of the PCS are identical with Fig. \ref{fig:fig2}. The $SiO_2$ box layer is 3$\mu$m. The background scattering is fitted by removing the air holes (dotted lines) and tuning the PCS layer index (dashed lines). \protect\hypertarget{fig8c}{(c)}\protect\hypertarget{fig8d}{(d)} Reflection spectra (black solid lines) and phase shift (red dashed lines) for the PCS on a silica substrate (c) and on the SOI structure (d). The insets are the $E_z$ field distribution of modes at 1537.2nm. }%
\label{fig:fig8}
\end{figure}

The Silicon-on-Insulator (SOI) technology is one of the most classical and important platforms to study the PCS characteristics for on-chip integration. Here, we consider an SOI configuration that has a 250nm silicon layer sitting on top of 3 $\mu m$  silicon dioxide layer and an infinite silicon substrate (Fig. \ref{fig:fig8}\hyperlink{fig8a}{(a)}). The SOI BOX layer performs as a Fabry-Perot (FP) cavity and modulates the background scattering matrix \textbf{C} in Eq. \eqref{eqn:eq3}. Different from the BBR where the high reflection is formed by the low-Q PCS modes, the reflection for the device on the SOI can be influenced by the wave destructive/constructive interference in the BOX layer. In this structure, the mode coupling strength to the transmission port is reduced by the indirect coupling via the FP mode. Due to the low confinement, the FP mode is a super-radiant mode with very broadband spectrum. Such two-resonance interaction can be ultimately unified as a one-resonance model \cite{RN1} governed by Eq. \eqref{eqn:eq3}. 
\par

With Ex-polarized excitation on this SOI structure, a resonant mode at 195.026 THz (1.537$\mu m$) can be observed that couples with the FP background and forms a Fano lineshape (black solid line, Fig. \ref{fig:fig8}\hyperlink{fig8b}{(b)}. To get the background scattering spectrum, the air holes on the PCS layer are removed to eliminate the interference of the PCS guide modes. The effective index shift of the device layer when removing the air holes also slightly changes the phase penetration depth \cite{RN6}, resulting in a spectral shift (red dotted line, Fig. \ref{fig:fig8}\hyperlink{fig8b}{(b)}). The background scattering is fitted by a uniform effective medium of the silicon layer with the same thickness as PCS. The effective index is simulated as 3.4535 (blue dashed line, Fig. \ref{fig:fig8}\hyperlink{fig8b}{(b)}), according to the relation $\varepsilon _{eff}=(1-f_1)\varepsilon _{Si}+f_1\varepsilon _{air}$ \ where the filling ratio is related to the area of the holes in one unit cell $f_1=\pi r^2/a^2$. The S-parameters for the background scattering can then be retrieved by this background fitting. To demonstrate the enhancement of reflection coupling by the SOI structure, we compare the reflection spectra and phase shift for the same PCS configuration on a silica substrate (Fig. \ref{fig:fig8}\hyperlink{fig8c}{(c)}) with the above-mentioned SOI structure (Fig. \ref{fig:fig8}\hyperlink{fig8d}{(d)}). The maximum phase shift is less than $\pi$ for the PCS on silica structure, while the PCS on SOI structure shows a complete $2\pi$ phase shift. This comparison shows that the FP mode that is confined between the silicon substrate and the PCS layer can contribute to the enhancement of reflection mode coupling and phase shift (i.e., $d_1$ and $\Gamma$). The corresponding $E_z$ field distribution of the modes on the two substrates shows a phase different, which induces the different phase shift in the reflection. 
\par

\begin{figure}[htbp]
\centering\includegraphics[width=8.6cm]{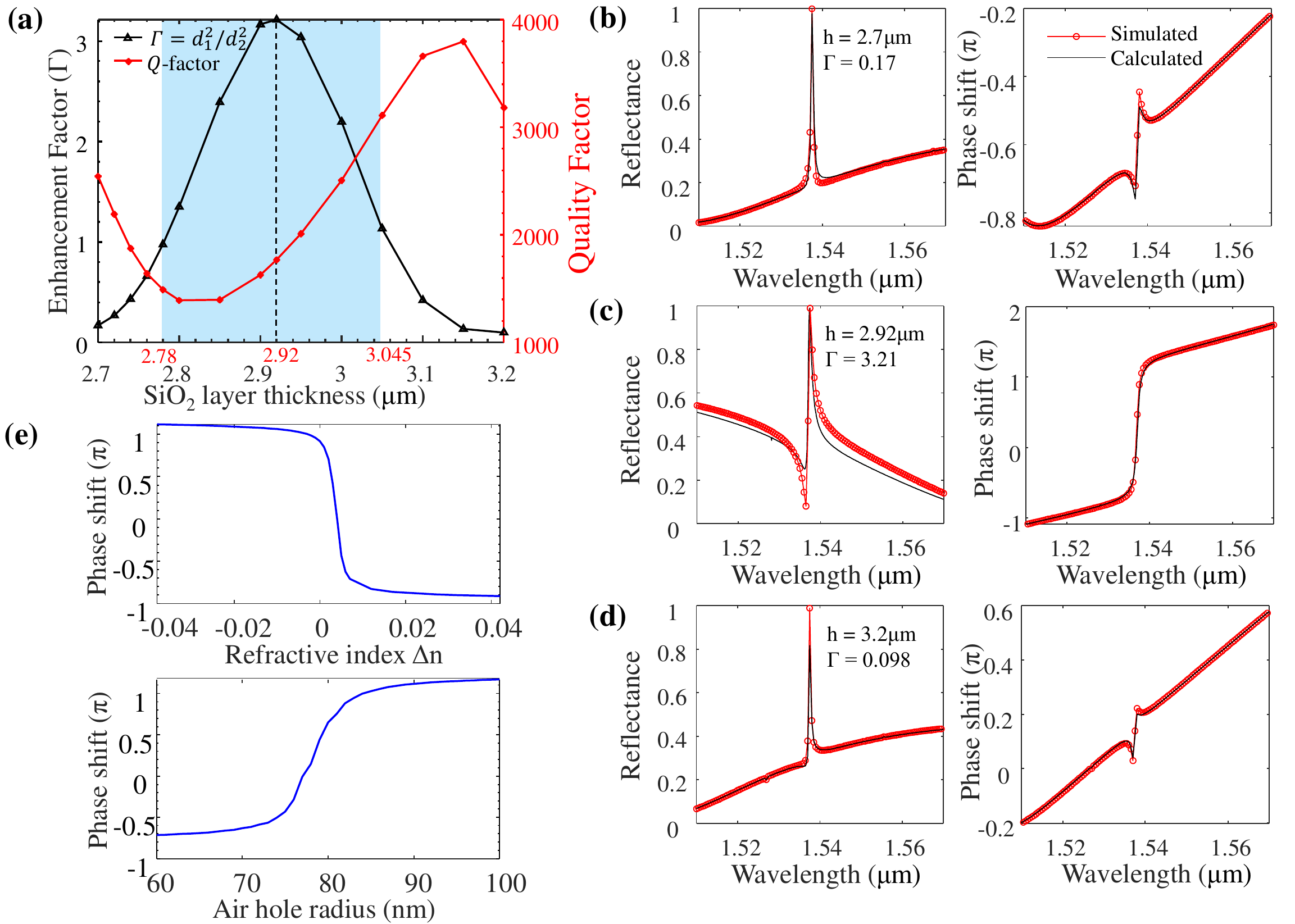}%
\caption{Reflection phase shift of the PCS on SOI structures. \protect\hypertarget{fig9a}{(a)} Enhancement factor and Q-factor with different $SiO_2$ layer thickness. The shaded region has enhanced coupling ($\Gamma > 1$) and thus $2\pi$ phase shift in the reflected wave. \protect\hypertarget{fig9b}{(b)}\protect\hypertarget{fig9c}{-}\protect\hypertarget{fig9d}{(d)} Reflection spectra and phase shift with $SiO_2$ thickness of $2.7\mu m$, $2.92\mu m$, and $3.2\mu m$. \protect\hypertarget{fig9e}{(e)} Near $2\pi$ phase shift control by index tuning (top panel) and air hole radius variation (bottom panel). The phase shift is probed with 195THz TEM incidence. }%
\label{fig:fig9}
\end{figure}

As a matter of fact, we will show that the interaction between the background FP mode and the PCS guided resonant mode can control the phase shift in the reflection wave. The enhancement factor ($\Gamma$) can be calculated by the coupled-mode equations with the fitted background scattering matrix as obtained in Fig.\ref{fig:fig8}\hyperlink{fig8b}{(b)}. Fig.\ref{fig:fig9}\hyperlink{fig9a}{(a)} shows the calculated enhancement factor ($\Gamma$) and Q-factor of the PCS mode with different dioxide layer thickness. The shaded area is the region that supports 2$\pi$ phase shift according to the condition $\Gamma>1$. The configuration of the PCS is the same as Fig. \ref{fig:fig8}. Sweeping the thickness of the $SiO_2$ layer from 2.7$\mu m$ to 3.2 $\mu$ m, we can find the maximum $\Gamma$ factor is 3.21 at thickness of $2.92\mu m$, and minimum $\Gamma$ factor is 0.098 for thickness of $3.2\mu m$. Reflection and phase shift spectra are simulated and calculated for 3 different $SiO_2$ layer thickness, as shown in Fig.\ref{fig:fig9}\hyperlink{fig9b}{(b)}$-$Fig.\ref{fig:fig9}\hyperlink{fig9d}{(d)}. At the maximum $\Gamma$, the reflection wave shows a complete $2\pi$ phase shift with a Fano reflection lineshape. While at the minimum $\Gamma$, the reflection lineshape has a reduced Fano asymmetric factor and the phase shift is smaller than $\pi$. 

\begin{figure}[htbp]
\center\includegraphics[width=8.6cm]{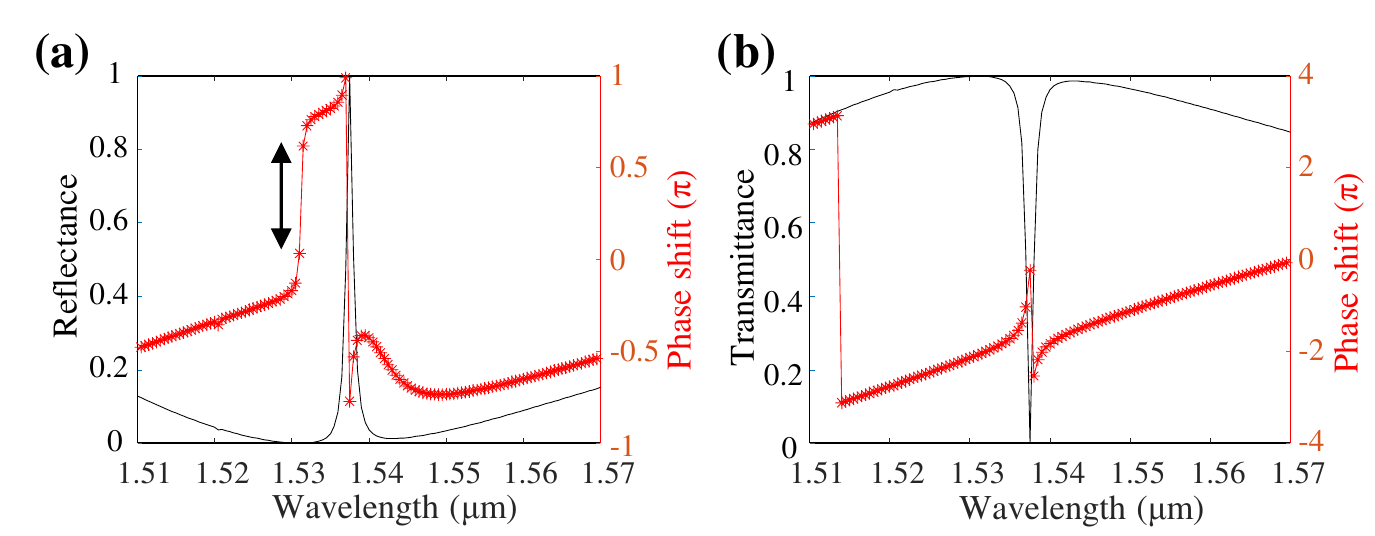}%
\caption{Reflection (left panel) and transmission (right panel) spectra for BOX layer height of h = 2.78$\mu $m where the enhancement factor $\Gamma{\approx}$1. The phase singularity is observed at around 1535nm.}%
\label{fig:fig10}
\end{figure}

The interaction with the FP mode can also create a singularity in the scattering matrix where the abrupt phase shift happens at $\Gamma\approx1$ (Fig.\ref{fig:fig10}). The phase shift control can be implemented by introducing parametric variations, such as lattice constant, air hole filling ratio, as well as PCS refractive index. The PCS layer index can be modulated by the temperature or dopant densities in silicon, depending on the operation speed of the applications. The index and reflection phase shift relation are shown in the top panel of Fig.\ref{fig:fig9}\hyperlink{fig9e}{(e)}) where a 0.05 index perturbation can trigger a near $2\pi$ phase shift for a 195THz incident wave. The bottom panel in Fig.\ref{fig:fig9}\hyperlink{fig9e}{(e)} shows the phase shift can also be modulated, under the monochromatic incidence, by the hole radius. Because the Q-factor with larger holes becomes lower, the phase shift variation is slower when expanding to the $2\pi$ phase range. The reduced sensitivity can provide more tolerance to the fabrication. In conclusion, the proposed heterostructures have the potential to enable phase sensing scheme and the spatial light modulation on the more compact PCS platform.

\section{Conclusion}
In this paper, we investigate a complete 2$\pi$ phase shift mechanism induced by the asymmetric mode couplings in a PCS heterostructure. The temporal coupled-mode calculation and the S-parameter simulation results show that the phase shift as well as lineshape function for the reflection wave can be efficiently controlled by such asymmetric designs. We study two scenarios that have enhanced mode coupling to the reflected wave. The first configuration designs a high-Q resonant mode residing in the high-reflection region in a broadband reflector formed by another two low-Q modes. Another configuration enhances the mode reflection coupling by modulating the background low-confinement FP mode in the SOI BOX layer. By designing the spectral distribution of the high-Q and low-Q guided modes with the PCS, extreme reflection lineshape with reduced Fano asymmetric factor can be achieved with a monotonous phase shift. Such configuration is an ideal candidate for the phase sensor design which features higher sensitivity compared to the traditional intensity peak-tracking sensors. A complete 2$\pi$ phase control is important for the phase sensors due to the monotonous phase mapping to the refractive index or lattice structure perturbations. The phase delay can be obtained by the self-interference experiments under single-frequency probing lasers. In this regard, the measurement setup for the sensor device can be simplified if the detection source and the PCS-based sensors are integrated on the chip. 
The single-layer PCS design provides a new route for integrating phase sensors with near-unity reflection, and monotonous phase-index relation. On the other hand, the PCS phase shift design method can be applied to coupled PCS structures ~\cite{RN89} for high-speed phase modulation as well as resonance lineshape control. The manipulation and control of reflection phase shift can be easily extended to the transmitted waves by reversing the asymmetricity of the heterostructure. The condition for the 2$\pi$ phase control is  \  $\left|d_1d_2\right|>\left|t\right|/\left|h\left(\omega
\right)\right|$, where t is the background transmission coefficient, $h\left(\omega \right)$ \ is the mode spectral lineshape, and $d_1,d_2$ \ are the coupling coefficient with the waves from the two ports. The ability to control the phase shift and mapping by the photonic crystals enables the promising functionalities of PCS-based metamaterial, phase sensors, beam steering, and more.

\begin{acknowledgments}
The authors would like to acknowledge the financial support from US Army Research Office grant (W911NF1910108)
\end{acknowledgments}

\bibliography{ms}

\newpage
\end{document}